\documentclass[prb,aps,superscriptaddress,floatfix,tightenlines,showpacs,twocolumn]{revtex4-2}
\usepackage{graphicx}
\usepackage{dcolumn}   
\usepackage{bm}        
\usepackage{amssymb}   
\usepackage{amsmath}
\usepackage{url}
\usepackage{layout}
\usepackage{color}
\usepackage{epsfig}
\usepackage{graphicx}
\usepackage{mathrsfs}\usepackage{bm}\usepackage{appendix}\usepackage{color}

\usepackage[thinlines]{easytable}
\usepackage{makecell}
\usepackage{tikz,pgf}
\usepackage{booktabs}
\usepackage{float}
\usepackage{hyperref}
\hypersetup{colorlinks,allcolors=blue}

\begin{document}

\title{Origin of the Diagonal Double-Stripe Spin-Density-Wave and Potential Superconductivity in Bulk La$_3$Ni$_2$O$_{7}$ at Ambient Pressure}

\author{Yu-Bo Liu}
\thanks{These three authors contributed equally to this work.}
\affiliation{School of Physics, Beijing Institute of Technology, Beijing 100081, China}

\author{Hongyi Sun}
\thanks{These three authors contributed equally to this work.}
\affiliation{Shenzhen Institute for Quantum Science and Engineering, Southern University of Science and Technology, Shenzhen 518055, China}
\affiliation{International Quantum Academy, Shenzhen 518048, China}

\author{Ming Zhang}
	\thanks{These three authors contributed equally to this work.}
	\affiliation{Zhejiang Key Laboratory of Quantum State Control and Optical Field Manipulation, Department of Physics, Zhejiang Sci-Tech University, 310018 Hangzhou, China}

\author{Qihang Liu}
\affiliation{Department of Physics, Southern University of Science and Technology, Shenzhen 518055, China}
\affiliation{Shenzhen Key Laboratory of Advanced Quantum Functional Materials and Devices, Southern University of Science and Technology, Shenzhen 518055, China}

\author{Wei-Qiang Chen}
\email{chenwq@sustech.edu.cn}
\affiliation{Department of Physics, Southern University of Science and Technology, Shenzhen 518055, China}
\affiliation{Shenzhen Key Laboratory of Advanced Quantum Functional Materials and Devices, Southern University of Science and Technology, Shenzhen 518055, China}

\author{Fan Yang}
\email{yangfan\_blg@bit.edu.cn}
\affiliation{School of Physics, Beijing Institute of Technology, Beijing 100081, China}

\begin{abstract}
\noindent The discovery of high-temperature superconductivity (SC) with $T_c\approx 80$ K in the pressurized La$_3$Ni$_2$O$_{7}$ has aroused great interests. Currently, due to technical difficulties, most experiments on La$_3$Ni$_2$O$_{7}$ can only be performed at ambient pressure (AP). Particularly, various experiments have revealed the presence of spin density wave (SDW) in the unidirectional diagonal double-stripe pattern with wave vector near $\pm(\pi/2,\pi/2)$  in La$_3$Ni$_2$O$_{7}$ at AP.  In this work, we employ first-principle calculations followed by the random phase approximation (RPA)-based study to clarify the origin of this special SDW pattern and the potential SC in La$_3$Ni$_2$O$_{7}$ at AP. Starting from our density-functional-theory band structure, we construct an eight-band bilayer tight-binding model using the Ni-$3d_{z^2}$ and $3d_{x^2-y^2}$ orbitals, which is equipped with the standard multi-orbital Hubbard interaction. Our RPA calculation reveals an SDW order driven by Fermi-surface nesting with wave vector $\bm{Q}\approx(0,\pm0.84\pi)$ in the folded Brillouin zone (BZ). From the view of the unfolded BZ, the wave vector turns to $\bm{Q}_0\approx\pm(0.58\pi,0.58\pi)$, which is near the one detected by various experiments. Further more, this SDW exhibits an interlayer antiferromagnetic order with a unidirectional diagonal double-stripe pattern, consistent with recent soft X-ray scattering experiment. This result suggests that the origin of the SDW order in La$_3$Ni$_2$O$_{7}$ at AP can be well understood in the itinerant picture as driven by Fermi surfaces nesting. In the aspect of SC, our RPA study yields an approximate $s^\pm$-wave spin-singlet pairing with $T_c$ much lower than that under high pressure. Further more, the $T_c$ can be strongly enhanced through hole doping, leading to possible high-temperature SC at AP. 
\end{abstract}\maketitle

\section{Introduction}

Recently, the experiments~\cite{Wang2023LNO} have reported that the bilayer nickelate La$_3$Ni$_2$O$_{7}$ exhibits superconductivity (SC) at 80 K under pressures above 14 GPa, which has been confirmed by subsequent experiments~\cite{YuanHQ2023LNO,Wang2023LNOb,wang2023LNOpoly,wang2023la2prnio7,zhang2023pressure,zhou2023evidence,wang2024bulk,li2024pressure}. Inspired by SC in La$_3$Ni$_2$O$_{7}$, efforts have been made to search for more nickelate superconductors in the Ruddlesden-Popper (RP) phase~\cite{Beznosikov2000,LACORRE1992495}, resulting in the discovery of SC in La$_4$Ni$_3$O$_{10}$ under high pressure (HP)~~\cite{zhu2024superconductivity,zhang2023superconductivity,huang2024signature,li2023trilayer}. These discoveries have rapidly sparked research into the electronic structure~\cite{yang2024orbital,wang2023structure,cui2023strain,sui2023rno,YaoDX2023,Dagotto2023,cao2023flat,zhang2023structural,huang2023impurity,geisler2023structural,rhodes2023structural,zhang2023la3ni2o6,yuan2023trilayer,li2024la3,geisler2024optical,li2017fermiology,wang2024non,Li2024ele,li2024distinguishing,zhou2024revealing,wang2024chemical,chen2024tri,Chen2024poly,Dong2024vis,Li2024design,puphal2024unconven}, the pairing mechanisms~\cite{WangQH2023,YangF2023,lechermann2023,Kuroki2023,HuJP2023,lu2023bilayertJ,oh2023type2,liao2023electron,qu2023bilayer,Yi_Feng2023,jiang2023high,zhang2023trends,qin2023high,tian2023correlation,jiang2023pressure,lu2023sc,kitamine2023,luo2023high,zhang2023strong,pan2023rno,sakakibara2023La4Ni3O10,lange2023mixedtj,yang2023strong,lange2023feshbach,kaneko2023pair,fan2023sc,wu2024deconfined,zhang2024prediction,zhang2024s,Yang2024effective,zhang2024electronic,yang2024decom,ryee2024quenched,Lu2024interplay,Ouyang2024absence}, and the correlated states~\cite{ZhangGM2023DMRG,Werner2023,shilenko2023correlated,WuWei2023charge,chen2023critical,ouyang2023hund,heier2023competing,wang2024electronic,botzel2024theory} of nickelate SC materials.
The nickelate superconductors, including the pressurized RP-phase La$_3$Ni$_2$O$_7$ and La$_4$Ni$_3$O$_{10}$ and previously synthesized infinite-layer nickelates Nd$_{1-x}$Sr$_x$NiO$_2$~\cite{li2019superconductivity,lee2023linear,nomura2022superconductivity,gu2022superconductivity},  have become a new platform for exploring high-temperature SC after cuprates~\cite{wu1987,lee2006,Schilling1993} and iron-based~\cite{kamihara2008,Ren2008} superconductors.

Currently, most experimental investigations on the RP-phase nickelates are carried out at ambient pressure (AP) because the HP circumstance strongly hinders the experimental detection of the physical properties of the materials. It is inspiring that, very recently, evidences of SC with $T_c$ beyond the McMillan limit (40 K) has been reported in the ultrathin La$_3$Ni$_2$O$_7$ film at AP~\cite{Ko2024signature,zhou2024ambient}, which will strongly push the development of this field. Therefore, studies on the RP-phase nickelates at AP are becoming more and more important. At AP, experiments on the bulk materials of the RP-phase nickelates have uncovered density-wave (DW) orders which compete with the SC, and such competition is key to unraveling the mechanism of SC.  Concretely, evidences of charge DW (CDW) and spin DW (SDW) are reported in La$_3$Ni$_2$O$_{7}$~\cite{Fukamachi2001,khasanov2024pressure,chen2024evidence,dan2024spin,chen2024electronic,Wang2022LNO,Kakoi2024,xie2024neutron,gupta2024anisotropic,feng2024unaltered,meng2024density,fan2024tunn,xu2024pressure,LI2024distinct,wu2001magnetic,liu2024electronic} and La$_4$Ni$_3$O$_{10}$~\cite{Fukamachi2001,zhang2020intertwined,xu2024origin,du2024correlated}, catching lots of theoretical interests~\cite{zhang2024emergent,Leonov2024Electronic,labollita2024assessing,ni2024first,Yi2024nature,jiang2024theory,chen2024non,zhang2024magnetic,lin2024magnetic,qin2024intertwined,Leonov2024Electronicc,labollita2023ele}. 

Here, we focus on the SDW discovered in bulk La$_3$Ni$_2$O$_{7}$ at AP through the muon spin relaxation~\cite{khasanov2024pressure,chen2024evidence}, the nuclear magnetic resonance (NMR)~\cite{Fukamachi2001,dan2024spin,Kakoi2024}, the neutron scattering~\cite{xie2024neutron} and the soft X-ray scattering~\cite{chen2024electronic,gupta2024anisotropic} measurements. Particularly, the soft X-ray scattering measurements\cite{chen2024electronic,gupta2024anisotropic} reveal the presence of an interlayer antiferromagnetic (AFM) SDW order with an intriguing unidirectional diagonal double-stripe pattern within each layer, with wave vector near $\pm(\pi/2,\pi/2)$. Note that the NMR\cite{dan2024spin} measurements suggest that such a stripy SDW order is a pure spin order, without coexisting CDW order. To fit the experimentally detected magnon dispersion of this SDW state with linear spin wave theory, an attempt based on the local moment description of the SDW order has been made, resulting in a Heisenberg spin model with various superexchange interactions, among which the interlayer superexchange interaction is more than an order of magnitude stronger than the intralayer ones~\cite{chen2024electronic}. However, such a Heisenberg spin model is problematic because if the interlayer superexchange interaction really overwhelms the intralayer one, the interlayer dimer state would be formed, instead of the SDW ordered state. The unreasonable fitting result of the SDW order based on the local moment description necessitates the itinerant description of the SDW order. 

Various theoretical attempts have been made to clarify the origin of the SDW order with the diagonal double-stripe pattern based on the itinerant picture. The random phase approximation (RPA) calculations~\cite{lin2024magnetic} adopting the tight-binding (TB) parameters for the HP phase yield an SDW wave vector distinct from the experimental one, suggesting that the magnetic structures under HP and AP are distinct, which necessitates the investigation of the SDW order in the AP circumstance. Note that the lattice structure and symmetry of La$_3$Ni$_2$O$_7$ at AP are different from those under HP, particularly with doubled unit cell. The unbiased density-functional-theory (DFT) calculations\cite{labollita2023ele,Mochizuki2018} obtained ferromagnetic or A-type AFM phase as the ground state, which is not consistent with the experiments. The Dynamic-mean-field-theory~\cite{wang2024electronic} or
DFT\cite{Leonov2024Electronic,labollita2024assessing,ni2024first} based calculations have provided the energies corresponding to a few candidate SDW patterns within a $4\times 4$ expanded magnetic unit cell associated with the assumed or calculated wave vector $\pm(\pi/2,\pi/2)$, resulting in the conclusion that the diagonal double-stripe pattern is really the energetically favored one. However, these calculations are somewhat biased as only a few candidate SDW patterns have been investigated among the infinite number of possible ones within the $4\times4$ magnetic unit cell. Note that an arbitrary $4\times4$ pattern will usually lead to not only a peak at $\pm(\pi/2,\pi/2)$ but also peaks at the higher harmonic components of this wave vectors, e.g. $(\pi/2,\pi)$ or $(\pi,\pi)$, while the experiments on La$_3$Ni$_2$O$_7$ at AP have only discovered the one $\pm(\pi/2,\pi/2)$~\cite{chen2024electronic,gupta2024anisotropic}. It is the purpose of the present study to reveal the origin of the special SDW pattern observed in experiments, including the wave vector near $\pm(\pi/2,\pi/2)$ and the unique unidirectional diagonal double-stripe pattern, from unbiased first-principle calculations without assuming any candidate patterns.

In this paper, we carry out an unbiased first-principle study on the properties of the SDW and the potential SC in La$_3$Ni$_2$O$_7$ at AP. In Sec.~\ref{sec:TB}, we start by constructing a bilayer eight-band TB model, obtained by Wannier fitting of our first-principle DFT+U band structure, using the Ni-$3d_{z^2}$ and $3d_{x^2-y^2}$ orbitals. Then, in Sec.~\ref{SDW}, after considering the standard multi-orbital Hubbard interaction, we perform a RPA based study on the properties of the SDW order of the system. Our result reveals an SDW order with wave vector $\bm{Q}\approx(0,\pm0.84\pi)$ in the folded Brillouin zone (BZ), which turns into $\bm{Q}_0\approx\pm(0.58\pi,0.58\pi)$ in the unfolded BZ, near the $\pm(\pi/2,\pi/2)$ detected by experiments. The real-space distribution of the magnetic moment of the SDW takes the unidirectional diagonal double-stripe pattern, with the moments in the two NiO$_2$ planes AFM aligned. Such an SDW pattern is well consistent with experiments. In Sec.~\ref{SDW}, we further study the potential SC at AP via the RPA approach, which shows that the $T_c$ at AP is much lower than that under HP, consistent with experiments. Furthermore, our RPA result predicts that the $T_c$ will be significantly enhanced with hole doping at AP. Sec.~\ref{DisCon} provides the discussion and conclusion.

\begin{figure}[htbp]
		\centering
		\includegraphics[width=0.45\textwidth]{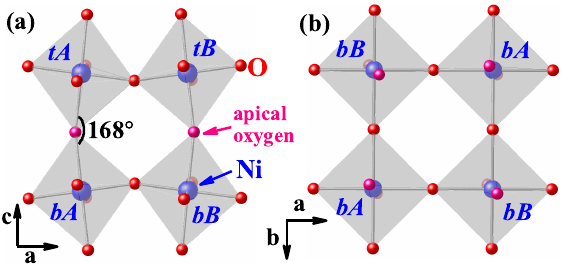}
		\caption{(color online) 
The schematic of the crystal structure of La$_3$Ni$_2$O$_7$ at AP, where red, pink, and blue spheres represent O, apical oxygen, and Ni atoms, respectively, and gray lines represent Ni-O bonds.  (a) is the side view of the bilayer crystal and (b) is the top view of the bottom layer crystal. The interlayer Ni-O-Ni bonding angle is 168$^\circ$, marked in (a).}
		\label{crystal}
\end{figure}
 
~~~~~~

\section{ DFT band structure and the microscopic model}
\label{sec:TB}

In La$_3$Ni$_2$O$_7$, each unit cell contains four $\text{NiO}$ layers\cite{Wang2023LNO}, including the weakly coupled isostructural upper and lower bilayers separated by a $\text{LaO}$ layer. The lattice structure of each bilayer is shown in Fig.~\ref{crystal}. In this structure, around each Ni atom, there are six oxygen atoms which form a standard octahedron. The two $\text{NiO}$ layers stacking along the c-axis are named as the top ($t$) and bottom ($b$) layer, respectively. Each Ni atom in the top layer is connected to a Ni atom in the bottom layer via an oxygen atom in the intermediate $\text{LaO}$ layer through the interlayer Ni-O-Ni bond. It is important to note that the interlayer Ni-O-Ni bonding angle at AP is 168$^\circ$, which is different from 180$^\circ$ under HP.  Such a difference leads to different lattice symmetries between the AP and HP phases: Under HP, the lattice structure belongs to the $Fmmm$ phase which possesses $C_4$ rotation symmetry and each unit cell contains one Ni atom in each layer. However, at AP, the lattice structure belongs to the $Amam$ phase, in which the octahedrons associate with the Ni atoms in the a-b plane alternately tilt toward two different directions, dividing the original Bravais lattice within each layer into two unequal sublattices labeled as $A$ and $B$ respectively, as shown in Fig.~\ref{crystal}(a-b). Meanwhile, the lattice lacks the $C_4$ rotation symmetry.


\begin{figure}[htbp]
		\centering
		\includegraphics[width=0.5\textwidth]{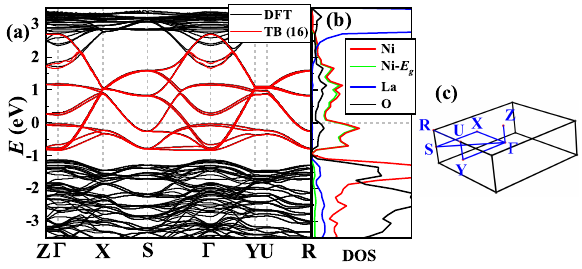}
		\caption{(color online) (a) DFT band structure (black lines) for La$_3$Ni$_2$O$_7$ at AP, in comparison with the sixteen-band TB band structure (red lines) obtained through Wannier fitting. Experimentally refined lattice constants are adopted in the DFT calculation. (b) The DOS near the Fermi energy, contributed by Ni (red), the E$_g$ orbitals of Ni (green) , La (blue) and O (black) atoms. (c)  Schematic of the 3D BZ. Several high symmetry points are marked in the BZ.}
		\label{dft}
\end{figure}

We have performed a DFT based calculation for the band structure of La$_3$Ni$_2$O$_7$ at AP, taking the experimentally detected lattice structure and the lattice constant\cite{VORONIN2001202}. The DFT calculations are performed based on the projector-augmented wave (PAW) pseudo-potentials with the exchange–correlation of the Perdew–Burke–Ernzerhof and the GGA + U approach, as implemented in the Vienna ab-initio Simulation Package (VASP)~\cite{dft1,dft2,dft3,Liechtenstein1995}. Starting from the $C_{mcm}$ crystal structure with experimentally measured lattice constants at AP~\cite{VORONIN2001202}, we fix the lattice constants and optimize the positions of atoms within the unit cell until the force acting on each atom is less than $10^{-3}$ eV$/\mathring{A}$. The cutoff energy is set to be $600$ eV, and the Kmesh is $19×\times19\times5$. To account for the correlation effects of $3d$ electrons in Ni atoms, an effective $U = 3.5$ eV is chosen, as reported in previous work~\cite{yang2024orbital}. Figure~\ref{dft}(a) shows our DFT band structure along the high symmetric lines in the BZ exhibited in Fig.~\ref{dft}(c). The density of state (DOS) contributed by different chemical elements is shown in Fig.~\ref{dft}(b), which shows that the low-energy DOS near the Fermi level is dominantly contributed by the Ni-$3d$-$E_g$ orbitals. Therefore, we project the Bloch electronic states obtained from DFT calculations onto the Ni-$3d_{z^2}$/$3d_{x^2-y^2}$ orbitals and construct a sixteen-band TB model Hamiltonian in the Wannier representation by using the Wannier90 Code~\cite{mostofi2008wannier90}. The sixteen-band TB band structure thus obtained is well consistent with the DFT one, as shown in Fig.~\ref{dft}(a) . 

In La$_3$Ni$_2$O$_7$ at AP, each unit cell contains four layers, with each layer containing two Ni atoms and each Ni atom containing two $3d$-$E_g$ orbitals, adding up to sixteen effective orbitals. It is important to note that the coupling between the upper bilayer and lower bilayer is very weak because the two bilayers are separated by a LaO layer, as also verified by the weak corresponding band split in the DFT band structure. Taking advantage of this character, we reasonably neglect the coupling between the upper bilayer and the lower one, and approximately take each bilayer as the unit cell to construct a 2D bilayer model. In this model, each unit cell only contains four $\text{Ni}$ atoms, labeled as $tA,tB,bA$ and $bB$ in Fig.~\ref{crystal}(a). The Hamiltonian of this eight-band TB model reads,
\begin{eqnarray}
		H_{\text{TB}}=\sum_{ij,\mu\nu,\sigma}t_{ij,\mu\nu}c^{\dagger}_{i\mu\sigma}c_{j\nu\sigma}+h.c.+\sum_{i\mu\sigma}\varepsilon_{\mu}c^{\dagger}_{i\mu\sigma}c_{i\mu\sigma}\label{equTB}.
	\end{eqnarray}
Here $i/j$ denotes the combined in-plane site, sublattice ($A$ or $B$) and layer ($t$ or $b$) index, $\mu/\nu$ labels orbital ($3d_{z^2}$ or $3d_{x^-y^2}$) and $\sigma$ labels spin. $\varepsilon_{\mu}$ is the on-site energy of orbital $\mu$. The hopping integrals $t_{ij,\mu\nu}$ of this model are extracted from the corresponding ones in the original sixteen-band model, which are provided in Tab.~S2 and Fig.~S2 in the Supplementary Materials (SM)~\cite{SM}. Similarly with La$_3$Ni$_2$O$_7$ under HP, the interlayer hoppings between the $d_{z^2}$ orbitals are the dominant ones.

\begin{figure}[htbp]
		\centering
		\includegraphics[width=0.45\textwidth]{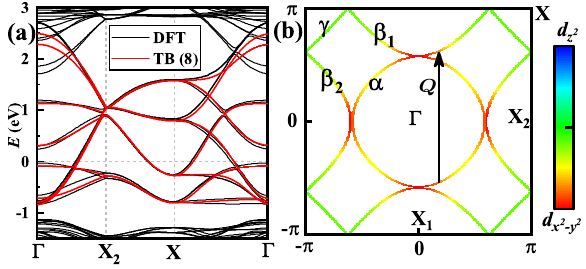}
		\caption{(color online) 
Band structure and Fermi surfaces (FSs) of La$_3$Ni$_2$O$_7$ at AP obtained from the eight-band TB-model Eq.~(\ref{equTB}). (a) The band structure of the eight-band TB-model (red line) along the high symmetry lines, compared with the DFT band structure (black line). (b) FSs in the folded BZ, marked by $\alpha,\beta_1,\beta_2$ and $\gamma$. The color in (b) indicates the orbital weight of $d_{x^2-y^2}$ and $d_{z^2}$. The FS-nesting vector is marked by $\bm{Q}$.}
		\label{tb}
\end{figure}

Figure~\ref{tb}(a) shows the band structure of the above eight-band TB model Eq.~(\ref{equTB}), in comparison with the DFT band structure. Obviously, the former has captured the essential characteristics of the latter near the Fermi energy. The FSs of the eight-band TB model are shown in Fig.~\ref{tb}(b), where the color indicates the orbital weight. There are four pockets, including an electron pocket $\alpha$ centered around the $\Gamma$ point, another small electron pocket $\gamma$ centered around the $M$ point, and two hole pockets $\beta_1$ and $\beta_2$ centered around the $X_1$ and $X_2$ points, respectively. Both orbital components are significantly involved on the FSs.

The band structure shown in Fig.~\ref{tb} illustrates two prominent features: (1) In comparison with the band structure under HP shown in Fig.~S1(c-d) in the $\mathbf{SM}$~\cite{SM} in which the bonding $d_{z^2}$- band crosses the Fermi energy to form a hole pocket $\gamma^{\prime}$, here at AP the bonding $d_{z^2}$- band shifts below the Fermi energy and the corresponding hole pocket vanishes. (2) The electron pocket $\alpha$ is well nested with the hole pocket $\beta_1$. Such FS nesting suggests the potential DW order at low temperatures, with the nesting vector $\bm{Q}$ as the wave vector. Note that as the unit cell at AP is doubled from that under HP, the BZ here is the inscribed square of the BZ under HP.

It is necessary to introduce the electron-electron interaction to investigate the correlated electronic states. We adopt the multi-orbital Hubbard interaction,
	\begin{align}\label{model}
		H_{\text{int}}&=U\sum_{i\mu}n_{i\mu\uparrow}n_{i\mu\downarrow}+
		(U-2J_H)\sum_{i,\sigma,\sigma^{\prime}}n_{i1\sigma}n_{i2\sigma^{\prime}} \nonumber\\
		&+J_{H}\sum_{i\sigma\sigma^{\prime}}\Big[c^{\dagger}_{i1\sigma}c^{\dagger}_{i2\sigma^{\prime}}c_{i1\sigma^{\prime}}c_{i2\sigma}+(c^{\dagger}_{i1\uparrow}c^{\dagger}_{i1\downarrow}c_{i2\downarrow}c_{i2\uparrow}+h.c.)\Big],
	\end{align}
where the first, second and third terms denote the intra-orbital Hubbard repulsion, the inter-orbital one and the Hund's rule coupling (plus the pair hopping), respectively. We fix $J_H = U/6$ for the subsequent calculations.

The total Hamiltonian adopted in our study is 
\begin{equation}\label{Hamiltonian}
H=H_{\text{TB}}+H_{\text{int}}.
\end{equation}
To solve this Hamiltonian, we adopt the RPA approach~\cite{takimoto2004strong,yada2005origin,kubo2007pairing,graser2009near,liu2013d+,zhang2022lifshitz,kuroki101unconventional}. In the RPA approach, by defining the spin susceptibility matrix $\chi^{(s)pq}_{st}(\bm{k},i\omega=0)$, the critical interaction strength $U_c \approx 1.2$ eV for SDW can be provided. For $U<U_c$, the finite spin susceptibility implies short-range spin fluctuations, which can mediate attractive interaction channels for SC. For $U>U_c$, the spin susceptibility diverges, suggesting the onset of SDW order. See $\mathbf{SM}$~\cite{SM} for details of the RPA approach.

~~~~~~

\section{\bf Properties of the SDW} \label{SDW}

\begin{figure}[htbp]
		\centering
		\includegraphics[width=0.48\textwidth]{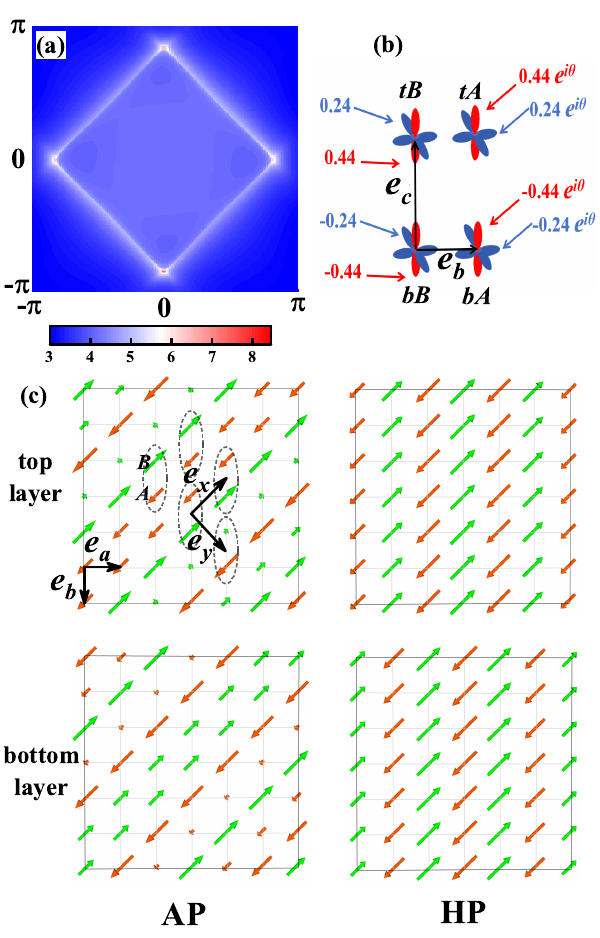}
		\caption{(color online) The wave vector and the pattern of SDW for La$_3$Ni$_2$O$_7$ at AP. (a) The distribution of the spin susceptibility $\chi^s(\bm{k})$ in the folded BZ for $U\approx 1$ eV. The maximal value of $\chi^s(\bm{k})$ just locates at $\bm{Q}\approx (0,\pm 0.84\pi)$. (b) The pattern of SDW within an unit cell, in which red (blue) orbital represents Ni-$3d_{z^2}$ ($3d_{x^2-y^2})$ orbital. The values are normalized. The difference between the SDW phases for the sublattice $A$ and $B$ within a unit cell is $\theta\approx 0.58\pi$. (c) Schematic SDW pattern on the whole lattice at AP (left panel) and under HP (right panel). In the left panel, each dashed oval represents a unit cell within a layer. The length and orientation/color of the arrows represent the magnitude and sign of the magnetic moment $m_{i\alpha}$, respectively. }
		\label{sdw}
\end{figure}

We study the properties of the SDW order in La$_3$Ni$_2$O$_{7}$ at AP via the RPA approach, including the wave vector and the distribution pattern of the SDW moment within a unit cell. To obtain the wave vector of the SDW order, we calculate the largest eigenvalue $\chi^{(s)}(\bm{k})$ of the spin susceptibility matrix $\chi^{(s)pq}_{st}(\bm{k}, i\omega_n=0)$. See $\mathbf{SM}$~\cite{SM} for more details of the spin susceptibility matrix $\chi^{(s)pq}_{st}(\bm{k}, i\omega_n)$. The $\chi^{(s)}(\bm{k})$ for $U=1$ eV is shown in Fig.~\ref{sdw}(a). Obviously, the largest $\chi^{(s)}$ is located near the momenta  $\bm{Q}=(0, \pm0.84\pi)$, which are precisely the nesting vectors between the $\alpha$ and $\beta_1$ Fermi pockets, as marked in Fig.~\ref{tb}(b). The second largest spin susceptibility is located at the momenta $\bm{Q}^{\prime}=(\pm 0.84\pi,0)$, corresponding to the nesting vector between the $\alpha$ and $\beta_2$ pockets. Note that the wave vectors $\bm{Q}$ and $\bm{Q}^{\prime}$ are not equivalent due to breaking of the $C_4$ symmetry in La$_3$Ni$_2$O$_7$ at AP. Our result yields that the spin susceptibilities $\chi^{(s)}(\bm{k}=\pm\bm{Q})$ are always the largest for any value of $U$ and $J_H$, suggesting that the wave vector of the SDW order in La$_3$Ni$_2$O$_{7}$ at AP should be $\pm\bm{Q}$.

The distribution pattern of the magnetic moment within a unit cell is determined by the eigenvector $\xi$ corresponding to the largest eigenvalue of the spin susceptibility matrix $\chi^{(s)}(\bm{Q}, i\omega_n=0)$ with $U\rightarrow U_c$, see $\mathbf{SM}$~\cite{SM} for more details. Our numerical results yield that the SDW order here is dominated by the on-site intra-orbital magnetism. Fig.~\ref{sdw}(b) shows the obtained distribution pattern, in which the values are normalized. Three features are obvious in this pattern. (1) For each orbital-site, the magnetic moment in the top layer is different from that in the bottom layer only by a minus sign, suggesting an interlayer AFM. (2) For each site in each layer, the ratio between the magnetic moment in the $3d_{x^2-y^2}$ orbital and that in the $3d_{z^2}$ orbital is fixed at about $0.24:0.44$, suggesting that both orbitals are significantly involved in the SDW order. (3) For each orbital and layer, while the SDW amplitudes of the $A$ and $B$ sites within any unit cell are the same, the difference between their SDW phases is $\theta_A-\theta_B=\theta\approx 0.58\pi$. It is interesting that this value is just very near half of the $y$-component of the wave vector $(0,2\pi)+\bm{Q}$.

In the following, we plot the distribution pattern of the SDW moment all over the lattice. Since the ratio between the magnetic moments of the two orbitals is fixed on any site, we can focus on the distribution pattern for any orbital, e.g. $3d_{z^2}$, as a representative in the following. In the SDW ordered state with wave vector $\bm{Q}$, the magnetic moment $m_{i\alpha}$ in the $i$-th unit cell in the sublattice $\alpha$ ($\alpha=A/B$) in, say the top layer, is given as,
\begin{eqnarray}
m_{i\alpha}&=&m_{0}\text{Re}\left[e^{i(\bm{Q}\cdot\bm{R}_i+\theta_\alpha)}\right]\nonumber\\&=&m_{0}\cos{(\bm{Q}\cdot\bm{R}_i+\theta_\alpha)},\label{s}
\end{eqnarray}
and that in the bottom layer differs from Eq.~(\ref{s}) by a minus sign. Note that in obtaining Eq.~(\ref{s}), we have used the fact that the SDW amplitudes in both sublattices are equal.  Here for convenience, we define
\begin{equation}
\bm{Q}\cdot\bm{R}_i+\theta_\alpha\equiv \theta_{i\alpha}\label{phase}
\end{equation}
as the phase of the SDW at the site in the $i$-th unit cell on the sublattice $\alpha$. The difference between the phases of the $A$ and $B$ sites within a unit cell is $\theta_A-\theta_B=\theta\approx0.58\pi$. Meanwhile, the wave vector of the SDW order is $\bm{Q}$ in the folded BZ, which is equally viewed as $(0,2\pi)+\bm{Q}=(Q_x,Q_y)\approx(0,1.16\pi)$. Note that we have $\theta\approx Q_y/2$. The distribution pattern obtained from Eq. (\ref{s}) is schematically illustrated in the left panel of Fig.~\ref{sdw}(c), which clearly exhibits a unidirectional diagonal double-stripe pattern. Here, each unit cell represented by a dashed oval contains the $A$ and $B$ sites, and the $\bm{e}_x$ and $\bm{e}_y$ are the unit vectors for this unit cell, while the $\bm{e}_a$ and $\bm{e}_b$ are the unit vectors for the original undoubled unit cell. In Fig.~\ref{sdw}(c), the length and orientation/color of the arrows represent the magnitude and sign of $m_{i\alpha}$, respectively.

In the left panel of Fig.~\ref{sdw}(c), along the $\bm{e}_x$-direction, $\theta_{i\alpha}$ remains constant because $Q_x=0$. Consequently, the magnetic moment remains constant along this direction,  leading to a diagonal stripe that extends along the $\bm{e}_x$-direction. Along the $\bm{e}_y$-direction, $\theta_{i\alpha}$ changes by $Q_y=1.16\pi$ with each translation by $\bm{e}_y$. Consequently, the magnetic moment nearly changes by a minus sign with each translation by $\bm{e}_y$, considering that $Q_y$ is near $\pi$. Note that each translation by $\bm{e}_y$ leads to passing across two stripes. Therefore, this distribution pattern just provides the unidirectional diagonal double-stripe pattern shown in the left panel in Fig.~\ref{sdw}(c). 

Then, let us investigate how the SDW phase $\theta_{i\alpha}$ changes with each translation by $\bm{e}_a$ or $\bm{e}_b$, so that to figure out its wave vector in the original unfolded BZ. With a translation by $\bm{e}_a$, if the translation is from sublattice $A$ to $B$, $\theta_{i\alpha}$ changes by 
\begin{eqnarray}
[(\bm{R}_i+\bm{e}_y)\cdot\bm{Q}+\theta_B]-(\bm{R}_i\cdot\bm{Q}+\theta_A)
=Q_y-\theta,
\end{eqnarray}
and if it is from $B$ to $A$, $\theta_{i\alpha}$ changes by
\begin{eqnarray}
[(\bm{R}_i+\bm{e}_x)\cdot\bm{Q}+\theta_A]-(\bm{R}_i\cdot\bm{Q}+\theta_B)=Q_x+\theta=\theta.
\end{eqnarray}
With the translation by $\bm{e}_b$, if the translation is from sublattice $A$ to $B$, the SDW phase $\theta_{i\alpha}$ changes by 
\begin{eqnarray}
&[&(\bm{R}_i+\bm{e}_y-\bm{e}_x)\cdot\bm{Q}+\theta_B]-(\bm{R}_i\cdot\bm{Q}+\theta_A)\nonumber\\
&=&Q_y-Q_x-\theta=Q_y-\theta,
\end{eqnarray}
and if it is from $B$ to $A$, $\theta_{i\alpha}$ changes by
\begin{eqnarray}
(\bm{R}_i\cdot\bm{Q}+\theta_A)-(\bm{R}_i\cdot\bm{Q}+\theta_B)
=\theta.
\end{eqnarray}
Interestingly, as $\theta\approx Q_y/2$, we have $Q_y-\theta\approx\theta\approx0.58\pi$.  Therefore, with each translation by $\bm{e}_a$ or $\bm{e}_b$, the SDW phase changes approximately $0.58\pi$, suggesting that the SDW wave vector in the unfolded BZ should be approximately $\bm{Q}_0=\pm(0.58\pi,0.58\pi)$.

The obtained SDW wave vector $\bm{Q}_0=\pm(0.58\pi,0.58\pi)$ in the unfolded BZ is close to the value $\pm(0.5\pi,0.5\pi)$ revealed by various experiments~\cite{dan2024spin,chen2024electronic,gupta2024anisotropic}, and the interlayer AFM SDW order with unidirectional diagonal stripe pattern shown in the left panel of Fig.~\ref{sdw}(c) is consistent with the experiment of soft X-ray scattering~\cite{chen2024electronic,gupta2024anisotropic}. Such consistency between our RPA results and the experimental observations suggests that the origin of the SDW order in  La$_3$Ni$_2$O$_{7}$ at AP can be well understood in the itinerant picture as being induced by FS-nesting. For comparison, the distribution of the magnetic moment under HP is schematically illustrated in the right panel of Fig.~\ref{sdw}(c), which exhibits a stripe pattern extending along the $\bm{e}_b$ direction with the wave vector $\bm{Q}_{1}\approx(0.9\pi,0)$\cite{YangF2023} in the unfolded BZ. Such an SDW order originates from the FS-nesting between the $\gamma^{\prime}$ and $\beta^{\prime}$ pockets in the unfolded BZ, as shown in Fig.~S1(d) in the $\mathbf{SM}$~\cite{SM}.

~~~~~~

\section{\bf Potential SC } \label{SC}
~~~~~~~~~~

Currently, no definite evidence of SC has been reported in bulk La$_3$Ni$_2$O$_{7}$ at AP. Here, we explore the potential SC in this system for the following two reasons. On the one hand, it cannot be excluded that SC with a lower $T_c$ can exist in the ideal pure material, which is nevertheless suppressed by ingredients such as oxygen deficiency and impurities in the samples currently synthesized. On the other hand, the SDW detected at AP can also be short-range order. In the framework of RPA~\cite{takimoto2004strong,yada2005origin,kubo2007pairing,graser2009near,liu2013d+,zhang2022lifshitz,kuroki101unconventional}, SC emerges when $U<U_c$, which is mediated by short-range spin fluctuations.  In RPA, the pairing nature is characterized by the pairing eigenvalue $\lambda$, which is related to the $T_c$ via $T_c\propto e^{-1/\lambda}$, and the pairing symmetry is determined by the eigenvector of the linearized gap equation corresponding to its largest eigenvalue $\lambda$. See $\mathbf{SM}$~\cite{SM} for more details on RPA treatment of SC.

\begin{figure}[htbp]
		\centering
		\includegraphics[width=0.45\textwidth]{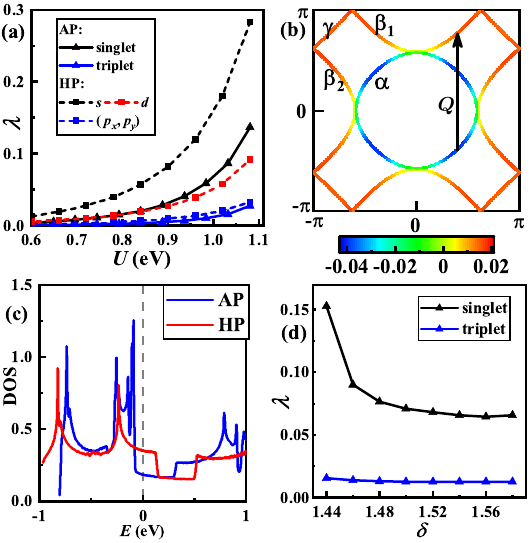}
		\caption{(color online) (a) The largest pairing eigenvalue $\lambda$ for La$_3$Ni$_2$O$_7$ at AP(solid lines) and HP(dashed lines) as function of $U$ for different pairing symmetries. (b) Distribution of the leading pairing gap function on the FS for $\text{U}=1$ eV. The FS-nesting vector is marked by $\bm{Q}$. (c) The DOS of La$_3$Ni$_2$O$_7$ under HP (red) and at AP (blue). (d) The largest pairing eigenvalue $\lambda$ at $\text{U}=1$ eV for the spin-singlet (black) and spin-triplet (blue) pairings as functions of the electron number $\delta$ per site. The results presented in (b) and (d) are obtained at AP.}
		\label{rpa}
\end{figure}

The $U$-dependent pairing eigenvalue $\lambda$ for La$_3$Ni$_2$O$_{7}$ at AP (solid lines) is shown in Fig.~\ref{rpa}(a), in comparison with that under HP (dashed lines). For La$_3$Ni$_2$O$_{7}$ under HP, whose band structure is given in $\mathbf{SM}$~\cite{SM}, the $D_4$ point group symmetry is present. In such a situation, the pairing symmetries include the $s$-wave, the $d$-wave, and the doubly-degenerate $(p_x,p_y)$-wave, and Fig.~\ref{rpa}(a) shows that the $s$-wave is the leading pairing symmetry. At AP, since the $D_4$ point group symmetry is absent, the pairing symmetries can only be distinguished as spin-singlet and spin-triplet. Fig.~\ref{rpa}(a) shows that the spin-singlet pairing is the leading pairing state. The distribution of the pairing gap function on the FS is shown in Fig.~\ref{rpa}(b). This pairing gap function is very close to an $s^{\pm}$-wave one, despite the lack of exact $C_4$ symmetry. The pockets $\alpha$ and $\beta_1$, connected by the nesting vector $\bm{Q}$, are distributed with the strongest pairing amplitude, with their pairing gap signs opposite.

Comparing the two sets of results for AP and HP in Fig.~\ref{rpa}(a), it can be observed that for any value of $U$, the leading $\lambda$ at AP is considerably smaller than that under HP. This result is consistent with experimental findings that high-$T_c$ SC has only been detected in La$_3$Ni$_2$O$_{7}$ under HP, and implies that SC with lower $T_c$ may exist at AP. In the following, we attempt to provide the physical explanation for the smaller $\lambda$ at AP, by comparing the band structures under HP and at AP. Under HP, The bonding $d_{z^2}$- band crosses the Fermi energy and forms a hole pocket $\gamma^{\prime}$, as shown in Fig.~S1(c-d) in the $\mathbf{SM}$~\cite{SM}. In contrast, at AP, the bonding $d_{z^2}$-band shifts below the Fermi energy, and consequently the corresponding hole pocket vanishes, as shown in Fig.~\ref{tb}. This significant difference affects SC in two aspects. Firstly, as shown in Fig.~\ref{rpa}(c), the DOS near the Fermi energy at AP is lower than that under HP. Secondly, the $\gamma^{\prime}$ pocket under HP provides another FS-nesting vector $\bm{Q}_1$ between the $\beta^{\prime}$ and $\gamma^{\prime}$ pockets, as shown in Fig.~S1(d) in the $\mathbf{SM}$~\cite{SM}, which contributes significantly to SC under HP\cite{WangQH2023,YangF2023}. In contrast, the absence of this pocket at AP, as shown in Fig.~\ref{tb}, considerably suppresses SC.


The impact of FS on the SC motivates us to investigate the effects of doping on SC. Our results are shown in Fig.~\ref{rpa}(d), where the $x$-axis represents the number $\delta$ of electrons per site, with $\delta=1.5$ corresponding to the undoping. The pairing eigenvalue $\lambda$ of the triplet pairing is always much smaller than that of the singlet pairing for all $\delta$, indicating that only singlet pairing is possible in this material. For the singlet pairing, the $\lambda$ remains almost unchanged under electron doping. Under hole doping, $\lambda$ first slowly increases with the enhancement of the doping until $\delta$ reaches $1.45$ where $\lambda$ promptly increases. The reason for the prompt enhancement of $\lambda$ for $\delta<1.45$ lies in the abrupt enhancement of the DOS shown in Fig.~\ref{rpa}(c), which originates from that the bonding $d_{z^2}$- band top touches the Fermi level at this doping. This result suggests that high-$T_c$ SC can be achieved in La$_3$Ni$_2$O$_{7}$ at AP through proper hole doping.

~~~~~~~~~~~~~

\section{{\bf Discussion and Conclusion}}\label{DisCon}
~~~~~~~~~~~~~~~~

The SDW wave vector $\bm{Q}_0\approx\pm(0.58\pi, 0.58\pi)$ obtained here slightly deviates from the $\pm(0.5\pi, 0.5\pi)$ detected experimentally. This might be caused by strong electron correlation neglected in our weak-coupling RPA approach. In the weak-coupling limit, the SDW wave vector would be exactly given as the FS-nesting vector, which is usually incommensurate. However, in the presence of strong electron interaction, some certain local SDW pattern would be energetically favored, which usually leads to formation of commensurate DW order. For realistic material, the SDW order might probably choose a commensurate wave vector near the FS-nesting vector, and consequently leads to the deviation between the SDW wave vector obtained here and that detected by experiments. 

In summary, we have systematically studied the SDW and potential SC for bulk La$_3$Ni$_2$O$_{7}$ at AP. By fitting the DFT band structure, the TB model at AP is obtained. Then we study the magnetic states through the RPA approach. Our results suggest the interlayer AFM SDW, hosting a special unidirectional diagonal double-stripe pattern with an in-plane wave vector $\bm{Q}_0\approx\pm(0.58\pi, 0.58\pi)$, which is qualitatively consistent with experiments. Such SDW state originates from the FS nesting. We have further studied the potential SC for La$_3$Ni$_2$O$_{7}$ at AP, obtaining an approximate $s^{\pm}$-wave SC with $T_c$ much lower than that under HP. Inspiringly, we find that hole doping can significantly enhance the $T_c$.
~~~~

\noindent{{\bf Acknowledgements}}

We are grateful to the discussions with Chen Lu. This work is supported by the NSFC under Grant Nos.12234016, 12074031, 12141402, and 12334002, Guangdong province (2020KCXTD001), Shenzhen Science and Technology Program under Grant No. RCJC20221008092722009. W.-Q. Chen is supported by the National Key R\&D Program of China (Grants No. 2024YFA1408101)), Guangdong Provincial Quantum Science Strategic Initiative Grand No. SZZX2401001, the SUSTech-NUS Joint Research Program, the Science, Technology and Innovation Commission of Shenzhen Municipality (No. ZDSYS20190902092905285), and Center for Computational Science and Engineering of Southern University of Science and Technology. Ming Zhang is supported by Zhejiang Provincial Natural Science Foundation of China under Grant No. ZCLQN25A0402.

~~~~

%

~~~~


\appendix
\renewcommand{\thetable}{S\arabic{table}}
\renewcommand{\thefigure}{S\arabic{figure}}
\appendix\setcounter{figure}{0}

~~~~

\end{document}